\documentclass[useAMS,usenatbib]{mn2e}

\usepackage{latexsym,graphicx,natbib}

\newcommand\simless{\mathbin{\lower 3pt\hbox
   {$\rlap{\raise 5pt\hbox{$\char'074$}}\mathchar"7218$}}} 
\newcommand\simgreat{\mathbin{\lower 3pt\hbox
   {$\rlap{\raise 5pt\hbox{$\char'076$}}\mathchar"7218$}}} 

\def\apgt{\ {\raise-.5ex\hbox{$\buildrel>\over\sim$}}\ }
\def\aplt{\ {\raise-.5ex\hbox{$\buildrel<\over\sim$}}\ }

\newcommand\kms{{\rm\,km\,s^{-1}}} 
\newcommand\kpc{{\rm\,kpc}} 
\newcommand\masyr{\rm\,mas\,{yr^{-1}}} 
\newcommand\msun{\rm\,M_\odot}

\newcommand\mbh{M_{\rm bh}}

\newcommand\he{HE\,0437--5439}
\newcommand\rmin{R_{\rm min}}
\newcommand\rpmin{R^{'}_{\rm min}}
\newcommand\rcore{r_{\rm c}}  
\newcommand\rhm{r_{\rm h}}  
\newcommand\trlx{T_{\rm rlx}}  
\newcommand\pcc{\rm\,pc^{-3}}  
       
\title[A hypervelocity star from the LMC]
      {A hypervelocity star from the Large Magellanic Cloud}

\author[Gualandris and Portegies Zwart]
       {
	 Alessia Gualandris $^{1}$\thanks{E-mail: alessiag@science.uva.nl} 
    	 and Simon Portegies Zwart$^{1}$\\
	 $^1$    Astronomical Institute 'Anton Pannekoek'
	 and Section Computational Science,
	 University of Amsterdam,
	 the Netherlands \\
       }

\begin{document}

\maketitle

\begin{abstract}
We study the acceleration of the star \he\, to hypervelocity and
discuss its possible origin in the Large Magellanic Cloud (LMC).  The
star has a radial velocity of 723$\kms$ and is located at a distance
of 61$\kpc$ from the Sun.  With a mass of about 8$\msun$, the travel
time from the Galactic centre is of about 100\,Myr, much longer than
its main sequence lifetime.  Given the relatively small distance to
the LMC (18$\kpc$), we consider it likely that \he\, originated in the
cloud rather than in the Galactic centre, like the other hypervelocity
stars.  The minimum ejection velocity required to travel from the LMC
to its current location within its lifetime is about 500$\kms$.
Such a high velocity can only be obtained in a dynamical encounter
with a massive black hole. We perform 3-body scattering simulations in
which a stellar binary encounters a massive black hole and find that a
black hole more massive than $10^3\msun$ is necessary to explain the
high velocity of \he. We look for possible parent clusters for \he\, and
find that NGC\,2100 and NGC\,2004 are young enough to host stars
coeval to \he\, and dense enough to produce an intermediate mass black
hole able to eject an 8$\msun$ star with hypervelocity.
\end{abstract}

\begin{keywords}
Stars: black-holes -- Stars: individual (HE\,0437-5439) --
Methods: N-body simulations -- Binaries:  dynamics.
\end{keywords}

\section{Introduction}
\label{sec:intro} 
Recent radial velocity measurements of young stars in the Galactic
halo has led to the discovery of a new class of stars, the so-called
{\it hypervelocity stars} (HVS), which travel with velocities
exceeding the escape speed of the Galaxy \citep{b05,b06}.  A total of
7 HVSs has been discovered so far, of which six are consistent with
an ejection from the centre of the Milky Way. 

The existence of a population of HVSs was first proposed by Hills
(1988), who considered hypervelocity ejections as a natural
consequence of galaxies hosting supermassive black holes (SMBHs).  The
interaction between a SMBH and a stellar binary can result in the
dynamical capture of one of the binary components at the expense of
the high velocity ejection of its companion star
\citep{h88,yt03,gps05}. Later an alternative was proposed by
\citet{bgp06}, who argue that an intermediate-mass black hole (IMBH)
can eject stars with extremely high velocities as it spirals in
towards the Galactic centre.  The latter mechanism tends to produce
HVSs in bursts which last a few Myr \citep{bgp06}.  Since the
reconstructed ejection times of the observed HVSs are roughly evenly
distributed over about 100\,Myr \citep{b06}, it seems likely that they
were ejected in a continuous process.  For 6 of the observed HVSs, in
fact, the travel time from the Galactic centre and the estimated age
are consistent and, by assuming a suitable proper motion, it is
possible to calculate a realistic trajectory.

For \he, however, an ejection from the Galactic centre appears
problematic as the time required to travel from the centre to its
observed location exceeds the maximum lifetime of the star.  \he\, is
a main-sequence star with spectral type B and a mass of about
8.4$\msun$ for a solar metallicity or of about 8$\msun$ for a $Z =
0.008$ metallicity, which corresponds to main-sequence lifetimes of 25
and 35\,Myr, respectively. Adopting a distance of 61$\kpc$ from the
Sun \citep{e05} and considering a total space velocity equal to the
radial velocity only, the travel time for a straight orbit from the
Galactic centre is of about 80\,Myr, much longer than the
main-sequence lifetime of the star.  The estimated proper motion for
which the orbit of the star comes within 10\,pc from the centre is of
about 0.55$\masyr$ \citep{e05}.  This results in a tangential velocity
of about 160$\kms$ and hence in a total space velocity of 740$\kms$,
which corresponds to a travel time of 100\,Myr for a realistic orbit.

We discard the possibility that \he\, is a blue straggler originated
in the Galactic centre since, even in the case of a merger between two
less massive stars, the merger product would not survive as an
8$\msun$ star for much longer than the main-sequence lifetime of such
a star.

Given the fact that \he\, is much closer to the LMC galaxy than to the
Milky Way, \citet{e05} suggest that \he\, might have been ejected from
the LMC.  Considering a total distance of 18$\kpc$ from the LMC
centre, the minimum ejection velocity required to travel to the
current position is about 500$\kms$.  (Here we adopted a travel time
of 35\,Myr, which is equal to the main-sequence lifetime in a low
metallicity environment.) The radial velocity of \he\, relative to the
LMC is 461$\kms$, and the required tangential velocity is then
160$\kms$, which in turns implies a proper motion of 1.9$\masyr$.  A
velocity of the order of 500$\kms$ can only be obtained in a dynamical
interaction with a massive black hole (see \citet{gps05}).

If \he\, was ejected by an IMBH and if it came from the LMC, the
natural consequence is that there must then be an IMBH in the LMC.  We
investigate this hypothesis by studying the dynamical ejection of an
8$\msun$ main-sequence star in an interaction with a massive
black hole. We simulate the encounter between a stellar binary and a
black hole to study the minimum mass of the black hole required to
accelerate one of the interacting stars to a velocity of at least
500$\kms$.  We then investigate where in the LMC such an interaction
could have taken place.

We find that a black hole of $\apgt 10^3\msun$ is required to explain
the velocity of \he.  Such an IMBH could form in a young and dense
star cluster \citep{p04} and would help driving the frequent strong
encounters with massive stars needed to make high-velocity ejections
likely \citep{2006MNRAS.372..467B}.  The IMBH must still be present in
a star cluster which contains stars as massive as \he. In addition,
the parent star cluster must have been massive enough and with a short
enough relaxation time to produce an IMBH \citep{p04}. The most likely
clusters are NGC\,2100 and NGC\,2004.

\section{Three-body scatterings with a massive black hole in the LMC}
\label{sec:scatter}

In this section we explore the hypothesis of a dynamical ejection from
the LMC by means of numerical simulations of three-body scatterings
with a massive black hole.  In \S\,\ref{sec:nb} we focus on
interactions in which a binary containing a young 8$\msun$
main-sequence star (representing \he) encounters a single black hole
of $10^2\msun$ to $10^4\msun$, whereas in \S\,\ref{sec:bbh} a single
8$\msun$ main-sequence star encounters a binary black hole.  In the
first case, an exchange interaction can lead to the high velocity
ejection of one of the binary components while in the latter case the
main-sequence star can be accelerated in a fly-by.

The simulations are carried out using the {\tt sigma3} package, which
is part of the STARLAB\footnote{\tt
http://www.manybody.org/manybody/starlab.html} software environment
\citep{mh96,p01}.

For each simulation we select the masses of the three stars, the
semi-major axis and eccentricity of the binary and the relative
velocity at infinity between the binary's centre of mass and the
single star.  The phase of the encountering binary is randomly drawn
from a uniform distribution \citep{hb83} and the orbital eccentricity
is taken randomly between circular and hyperbolic from the thermal
distribution.  The impact parameter $b$ is randomised according to
$b^2$ in the range $[0-b_{\rm max}]$ to guarantee that the probability
distribution is homogeneously sampled. The maximum value $b_{\rm max}$
is determined automatically for each set of experiments (see
\citet{gpe04} for a description).  Energy conservation is typically
better than one part in $10^6$ and, in case the error exceeds
$10^{-5}$, the encounter is rejected.  The accuracy in the integrator
is chosen in such a way that at most 5\% of the encounters are
rejected. During the simulations we allow for physical collisions when
the distance between any two stars is smaller than the sum of their
radii.  The black hole is assumed to be a point mass, while for the
stars we adopt zero-age main-sequence radii, taken from
\citet{1989ApJ...347..998E}.

\subsection{Encounters between a stellar binary and a massive black hole}
\label{sec:nb}

We consider encounters between a binary consisting of two
main-sequence stars with masses $m_1$ and $m_2$ and a single black
hole with mass $\mbh$.  The mass of one binary component is fixed to
8$\msun$ while the mass of the companion star has values of 2$\msun$,
4$\msun$, 8$\msun$ and 16$\msun$.  For each choice of stellar masses,
the semi-major axis $a$ is varied between a minimum value, which is
set by the physical radii of the stars so that the two components do
not touch at the first pericentre passage, and a maximum of 1\,AU.  We
consider black holes of masses $10^2\msun$, $10^3$ and $10^4\msun$.
The relative velocity at infinity between the black hole and the
binary's centre of mass is set equal to 20$\kms$.

\begin{figure}
\begin{center}
\includegraphics[width=8cm]{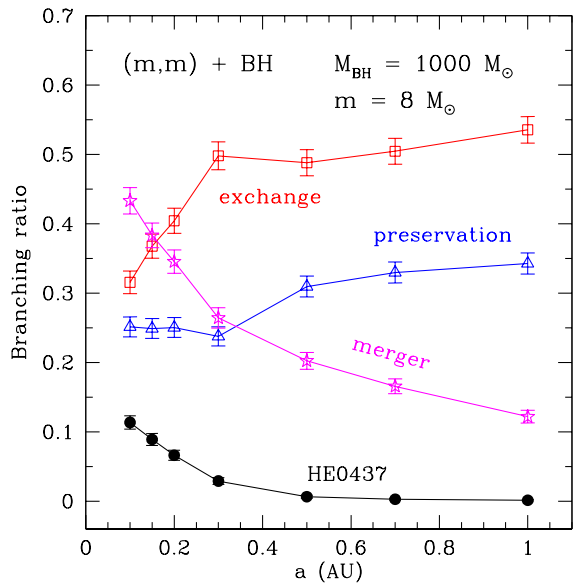}
\end{center}
\caption{Branching ratio for the outcome of encounters between a
  stellar binary and a single $10^3\msun$ black hole as a function of
  the initial binary semi-major axis.  The two binary components are
  assumed to be 8$\msun$ main-sequence stars.  The different outcomes
  are: merger (stars), preservation (triangles) and exchange
  (squares).  The subset of exchanges with a high velocity escaper
  ($V_{\infty} \ge 500\kms$) is indicated with bullets.  The error
  bars represent the formal ($1\sigma$) Poissonian uncertainty of the
  measurement.}
\label{fig:branch}
\end{figure}
In Fig.\,\ref{fig:branch} we present the probability of different
outcomes (branching ratios) in the simulations of an encounter between
an equal mass binary ($m_1 = m_2 = 8\msun$) and a $10^3\msun$ black
hole.  For each of the initial semi-major axes we perform a total of
1500 scattering experiments, which result either in a fly-by, an
exchange or a merger.  Mergers occur preferentially in the case of
tight binaries, leaving place to preservations and exchanges for wider
binaries.  In an exchange interaction one of the main-sequence stars
is captured by the black hole while the other star is ejected,
possibly with high velocity. If the velocity exceeds 500$\kms$, we
regard the star as a possible candidate for \he.  This occurs in about
10\% or less of all encounters for semi-major axes in the range
$0.1-1.0\,\rm AU$.

\begin{figure}
\begin{center}
\includegraphics[width=8cm]{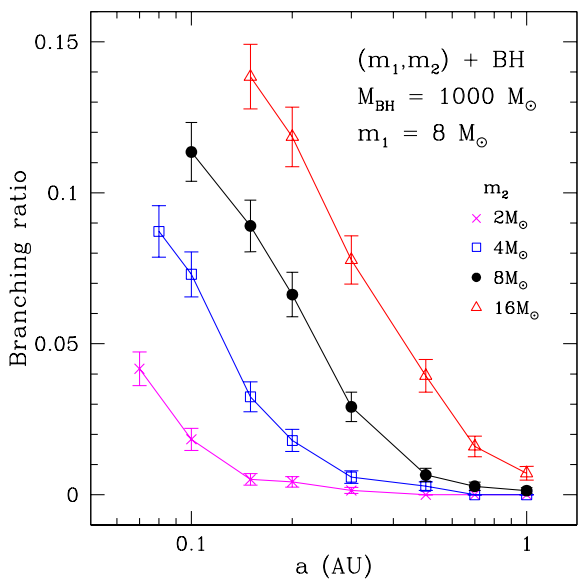}
\end{center}
\caption{Fraction of exchange encounters between a stellar binary and
         a $10^3\msun$ black hole that results in the ejection of a
         main-sequence stars with a velocity $\apgt 500\kms$.  The
         bullets indicate the relative probability for an incoming
         binary with two 8$\msun$ stars (see
         Fig.\,\ref{fig:branch}). The other curves are computed with
         different companion masses to an 8$\msun$ star: 2$\msun$
         (crosses), 4$\msun$ (squares), 16$\msun$ (triangles).  In the
         latter case, the ejected star is the secondary.}
\label{fig:branchHVS}
\end{figure}
In Fig.\,\ref{fig:branchHVS} we show the branching ratios for high
velocity ejections for binaries with a 8$\msun$ star and a companion
with mass in the range $2.0-16.0\msun$. While the 8$\msun$
main-sequence star is ejected, the companion star is captured by the
IMBH.  As in the previous plot, the black hole mass is of $10^3\msun$.
Exchange encounters with fast escapers are more likely in the case of
binaries with larger companion masses and/or shorter orbital periods.

Analytical estimates by \citet{yt03} for the case of tidal breakup 
predict an ejection velocity at infinity
\begin{equation}
\label{eq:tidal}
V_{\infty} = v^{'}\left(\frac{\mbh}{m}\right)^{1/6}\left(\frac{0.1\,\rm AU}{a}\frac{m}{1\msun}\right)^{1/2}\,,
\end{equation} 
where $v^{'}$ is a function of the distance of closest approach
$\rmin$. 
If we define a dimensionless closest approach parameter
$\rpmin = \frac{\rmin}{a} \left( \frac{\mbh}{m}\right)^{-1/3}\,,$
$v^{'}$ varies in the range 130--160$\kms$ for $\rpmin = 0-1$.
For $m=8\msun$ and $a=0.1\,\rm AU$ Eq.\,\ref{eq:tidal} yields
$v_{\infty} \approx 560\kms$ for $\mbh = 10^2\msun$; a 100$\msun$
IMBH would in principle be sufficient to explain the origin of \he.

\begin{figure*}
\begin{center}
\includegraphics[width=5.5cm]{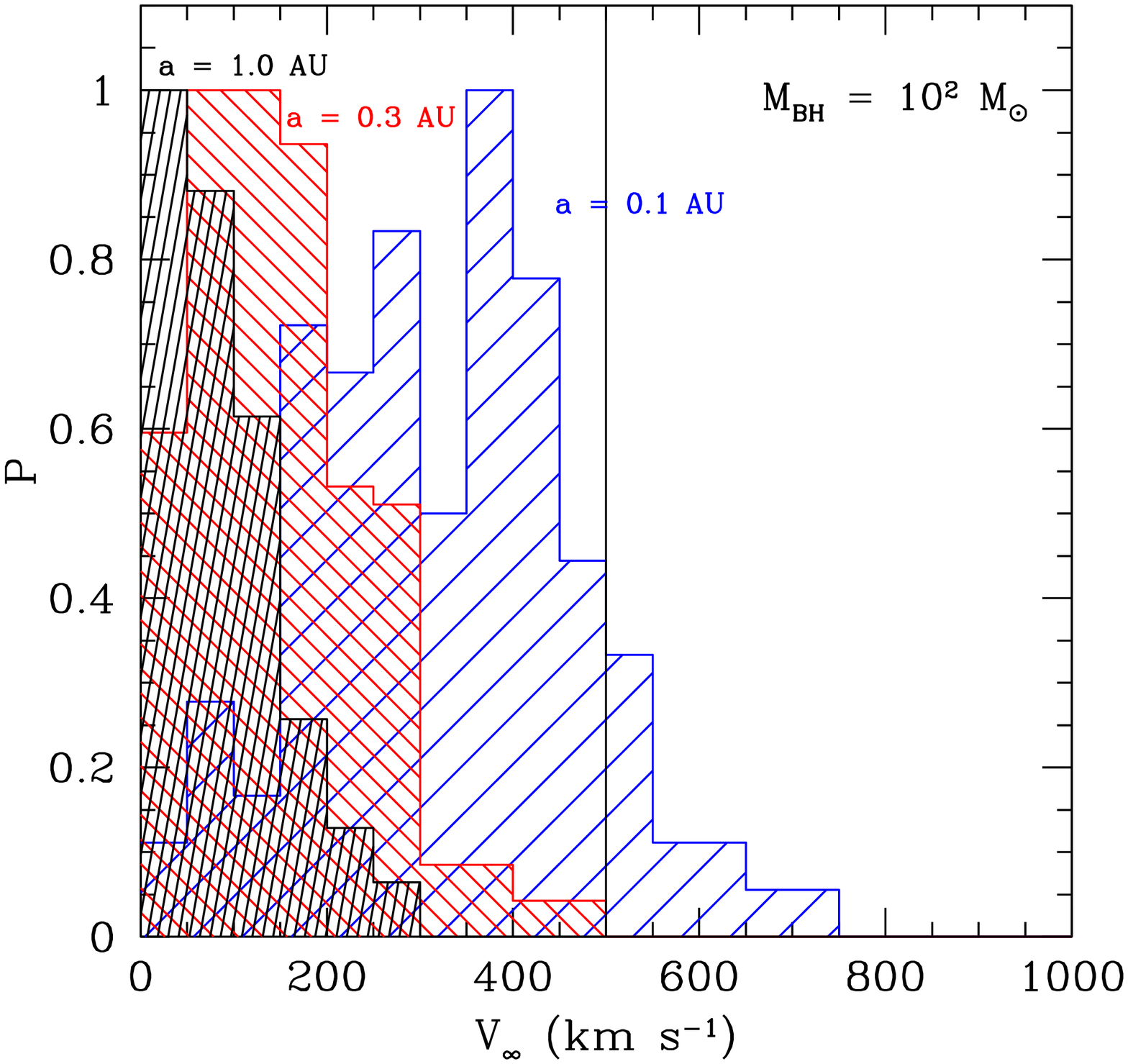}
\includegraphics[width=5.5cm]{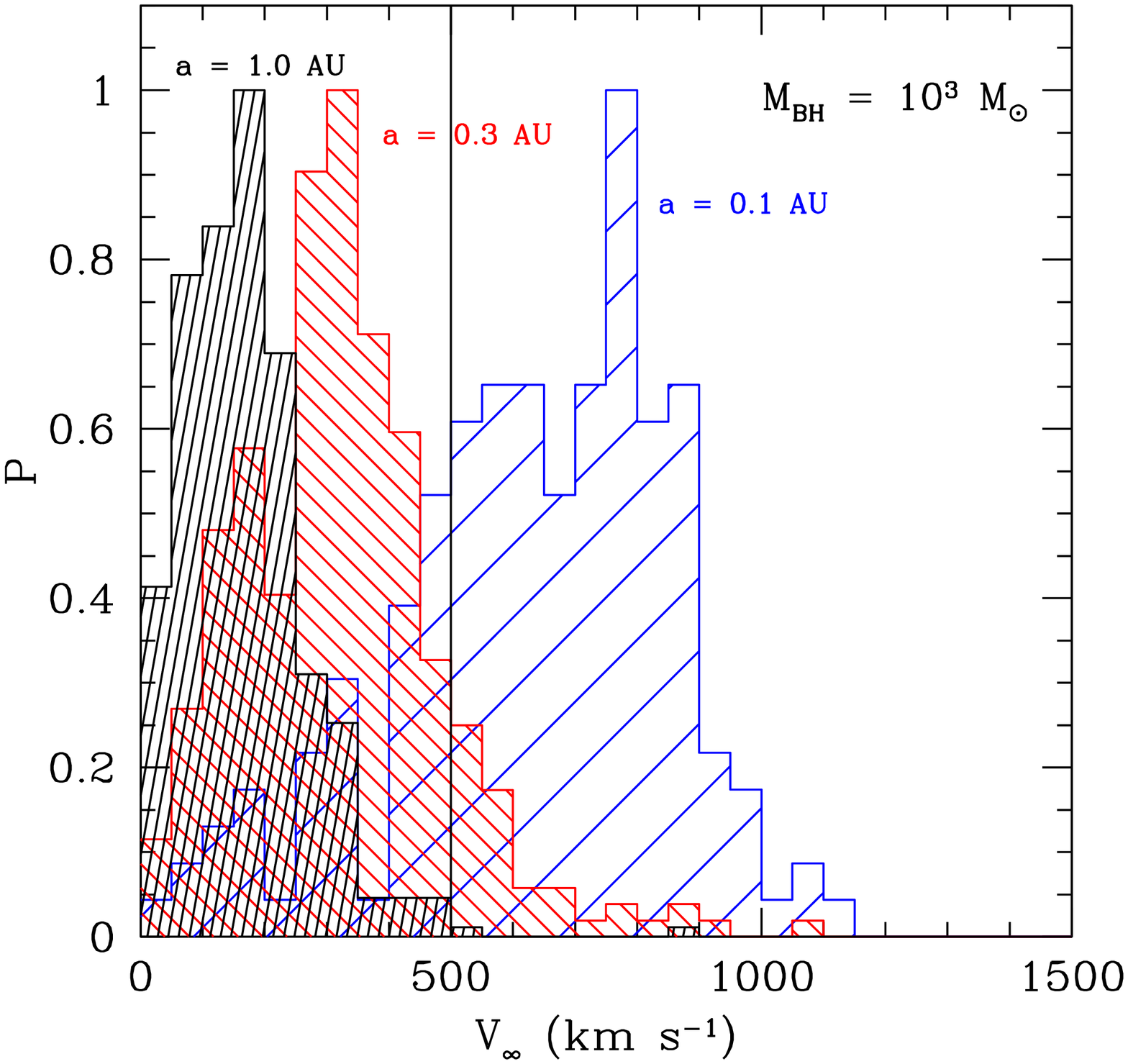}
\includegraphics[width=5.5cm]{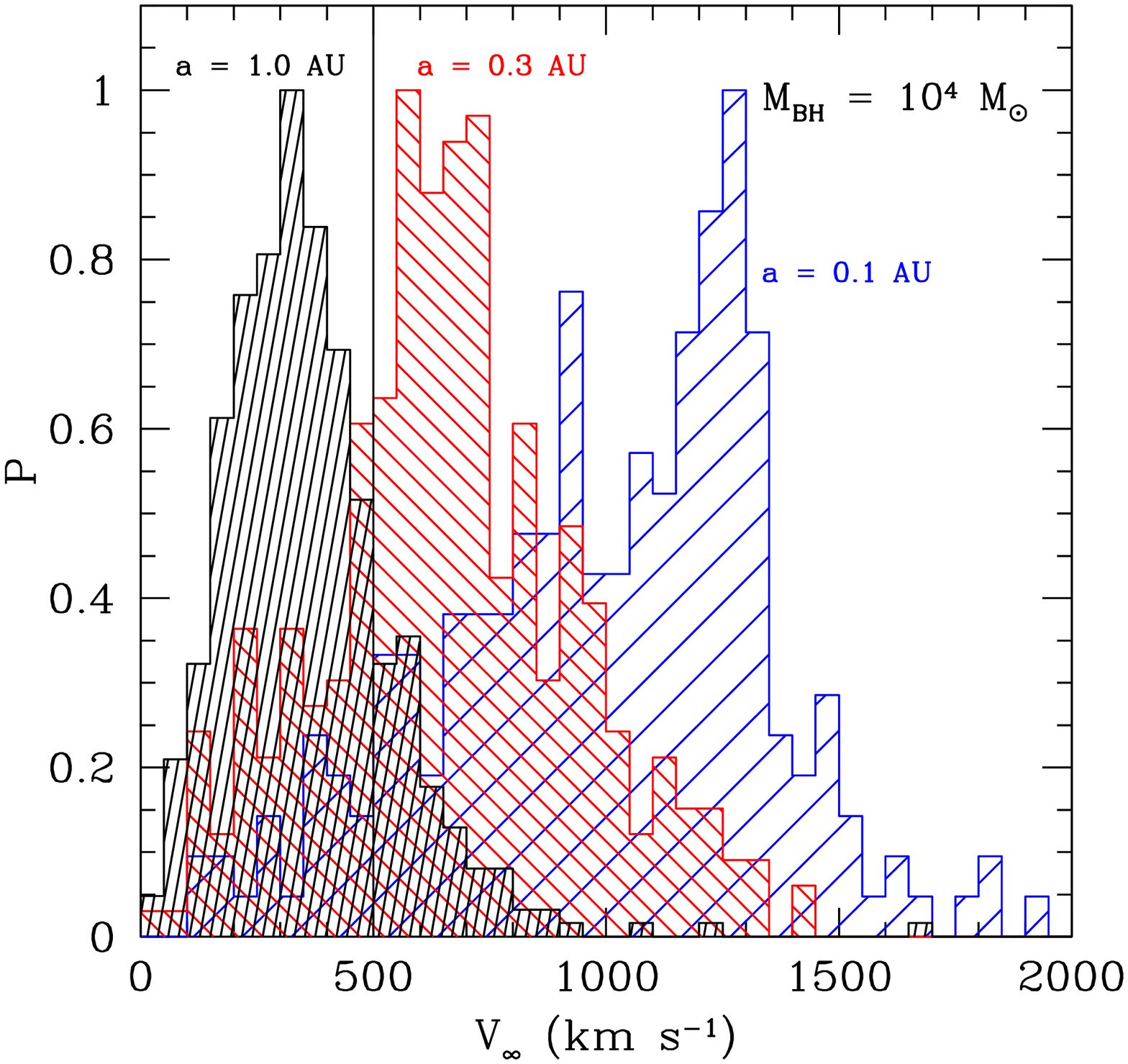}
\end{center}
\caption{Velocity distributions for different values of the black hole
         mass $\mbh=10^2\msun$ (left), $10^3\msun$ (middle),
         $10^4\msun$ (right), and of the initial semi-major
         axis $a$ = 0.1, 0.3, 1.0\,AU.}
\label{fig:veld}
\end{figure*}
In Fig.\,\ref{fig:veld} we show the velocity distributions of the
escaping single star in the case of equal mass binaries ($m_1 = m_2 =
8\msun$). The different panels refer to different values of the black
hole mass, from $10^2\msun$ (left) to $10^3\msun$ (middle) and
$10^4\msun$ (right). In each panel, the different distributions refer
to three different values of the initial semi-major axis.  While it
appears possible to reach the required recoil velocity in the case of
$\mbh=10^2\msun$, the smallest possible semi-major axis is needed
($a=0.1\,\rm AU$) for this to happen, and only in about 10\% of
exchange encounters, which corresponds to about 3\% of all simulated
scatterings.  In the case of the $10^4\msun$ black hole it is possible
to achieve such high velocities even for rather wide binaries. Based
on these data we conclude that a 100$\msun$ black hole is unlikely to
have resulted in the ejection of \he\, and a black hole mass $\apgt
10^3\msun$ is favoured for typical values of $a$.

We note that the velocity distributions are not sensitive to the
initial velocity between the binary and the single star as the total
energy of the system is dominated by the binding energy of the binary
rather than by the kinetic energy of the incoming star. Additional
experiments performed with initial velocities of $5\kms$ and $10\kms$
showed no appreciable difference in the ejection velocities.

We also note that for detached binaries the binary tidal radius is
always larger than the tidal radius of the single stars and therefore
tidal breakup and ejection occur before the stars can be disrupted by
the black hole. For (near) contact binaries, instead, it is possible
to disrupt a star before an ejection can take place.  This process
does not depend on the black hole mass but on the size of the binary
compared to the size of the stars. Our simulations only include
detached binaries as the proper treatment of contact binaries would
require hydrodynamical simulations.

The analytical estimates obtained with Eq.\,\ref{eq:tidal} are
considerably larger than the results of our simulations (see
Fig.\,\ref{fig:veld}).  This discrepancy is probably caused by the
assumption of \citet{yt03} that the binary is disrupted well within
the tidal radius, i.e. $\rpmin \aplt 1$, whereas in 10\% to 20\% of
our simulated encounters $\rpmin > 1$.  For example, for $\rpmin
\simeq 3-4$, a capture can occur which eventually leads to the
ejection of one star.  Since the ejection velocity depends largely on
the Keplerian velocity of the binary at the position of tidal breakup,
we expect a dependence of the final velocity of escapers on the
parameter $\rpmin$.
\begin{figure}
\begin{center}
\includegraphics[width=8cm]{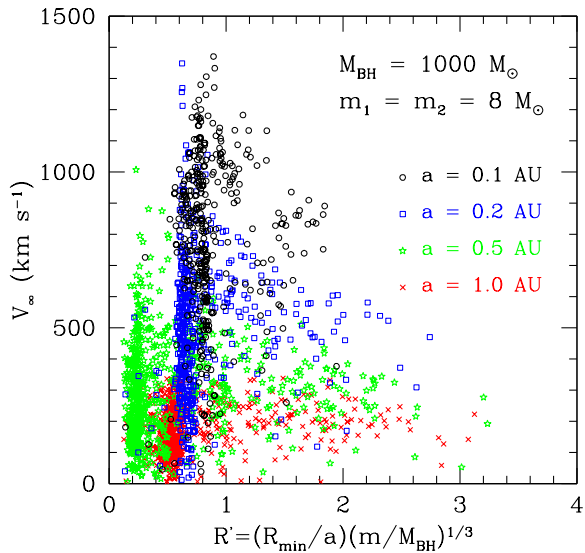}
\end{center}
\caption{Velocity of the ejected star after an exchange interaction
with a $10^3\msun$ black hole as a function of the distance of closest
approach to the black hole.  The different symbols indicate different
initial values of the binary semi-major axis.}
\label{fig:rmvel}
\end{figure}
Fig.\,\ref{fig:rmvel} shows the velocity at infinity of the ejected
star versus the dimensionless distance of closest approach.  In this
example we consider equal mass binaries with $m = 8\msun$ and a black
hole of mass $10^3\msun$.  
The figure shows that the highest ejection velocities are preferentially
achieved in close encounters. 
The neglect of relatively wide encounters by \citet{yt03} might
therefore be the origin of the apparent discrepancy in the velocities.
The agreement between our numerical simulations and the analytical
estimates of \citet{yt03} improves for higher black hole masses, and
is very good for encounters with a SMBH \citep[see][]{gps05}.

\begin{figure}
\begin{center}
\includegraphics[width=8cm]{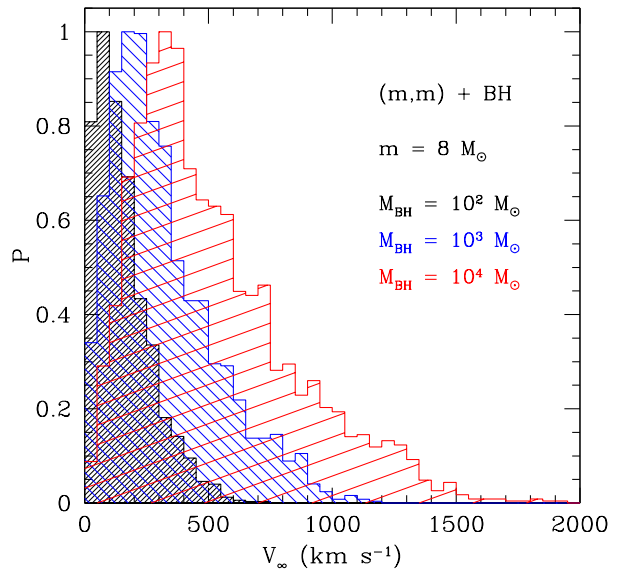}
\end{center}
\caption{Velocity distribution for the ejected star after an
         interaction between an equal mass ($m=8\msun$) binary and a
         black hole of $10^2\msun$ (black), $10^3\msun$ (blue)and
         $10^4\msun$ (red).  These velocity distributions are
         integrated over the entire range of orbital separations for
         the initial binary.}
\label{fig:all3}
\end{figure}

In our systematic study of the effect of the initial semi-major axis
of the interacting binary we adopted a homogeneous sampling in $\log
a$. Furthermore, the number of scattering experiments performed per
initial selection of $a$ are weighted with equal cross section. If the
distribution of orbital separations in a star cluster is flat in $\log
a$, like in the case of young star clusters
\citep{2005A&A...430..137K}, we can superpose the results of these
experiments in order to acquire a total velocity distribution of the
ejected star.  In Fig.\,\ref{fig:all3} we present this superposed
velocity distribution for a binary consisting of two 8$\msun$
stars. The three histograms in this figure give the velocity
distribution for a 100$\msun$ (left), $10^3\msun$ (middle) and
$10^4\msun$ (right) black hole.

\subsection{Encounters between a single star and a binary black hole}
\label{sec:bbh}

We now consider encounters between a single star and a binary black
hole.  In a recent study \citet{2006ApJ...640L..39G} argue that it is
possible to form a black hole binary as a result of two collision
runaways in a star cluster. The binary eventually merges due to the
emission of gravitational wave radiation \citep{peters64} but, before
coalescing, the two black holes experience frequent interactions with
other cluster stars, possibly accelerating them to larger velocities.

We simulate these encounters and study the probability of high
velocity ejections as a function of the binary parameters: the masses
of the two black holes $M_{\rm Bh1}$ and $M_{\rm Bh2}$ and the initial
semi-major axis $a$.  The mass of the single star is fixed to
8$\msun$, in accordance with the observed mass of \he. We consider
black hole masses of $10^2\msun$, $10^3\msun$ and $10^4\msun$ in both
equal mass and unequal mass binaries.  For each choice of black hole
masses, the semi-major axis is varied between a minimum value of 1\,AU
and a maximum value of 1000\,AU. The minimum value is chosen in order
to guarantee that the binary remains unaffected by the emission of
gravitational wave radiation during the period over which the
encounter takes place\footnote{Two $10^3\msun$\, black holes in a
1\,AU orbit will merge due to the emission of gravitational waves in
about 24\,Myr.}.

Except for a few mergers (about 1-2\% of all cases), the vast majority
of the encounters result in a fly-by.  The velocity at infinity of the
escaping star depends sensitively on the impact parameter of the
encounter.  Only in encounters for which the impact parameter is
comparable to the binary separation are high velocity ejections
realized. Most of these encounters are resonant, allowing the incoming
star to come as close to one the black holes as a fraction of the
orbital separation. Only a few \% of the encounters result in ejection
velocities as high as 500$\kms$ and we argue that this mechanism is
unlikely to have produced the observed velocity of \he.

\section{An IMBH in the LMC?}
The analysis carried out in the previous sections suggest the
intriguing idea that \he\, was ejected by an encounter with an IMBH of
$\apgt 10^3\msun$ in the LMC. Such an IMBH would most likely be found
in a young dense cluster containing stars coeval to \he.  Recently
\citet{mg2003} measured the structural parameters for 53 LMC clusters.
A total of nine clusters in this database are younger than 35\,Myr.
Three of these clusters, NGC\,1818, NGC\,1847 and NGC\,1850, have the
appropriate age to be the host of \he\, but their current densities
are much smaller than what is required to produce an IMBH
\citep{p04,2002MNRAS.330..232M}.  Of the six remaining clusters,
three (NGC\,1805, NGC\,1984, NGC\,2011) have a mass below 5000$\msun$
and are therefore unlikely to have produced an IMBH.  Also, the
presence of an IMBH in these clusters would have been noticed easily
\citep{2005ApJ...620..238B}.

\begin{table}
\caption{ List of young (age $\aplt 35$\,Myr) LMC clusters
with an estimated initial relaxation time smaller than 100\,Myr.  The
columns give the name of the cluster, the mass, the core radius, the
ratio of the core radius to the half-mass radius, the age, the current
two-body relaxation time, the initial two-body relaxation time
(computed using Eq.\,1 of \citet{2006astro.ph.10659P} with $\kappa =
0.1$) and the black hole mass based on \citet{hh06}.  For estimating
the relaxation time we adopted a mean mass $\langle m \rangle = 1$ and
a Coulomb logarithm parameter $\gamma = 0.4$ for the lower limit, and
$\langle m \rangle = 0.65$ and $\gamma = 0.01$ for the upper limit.
The core and half-mass radius for R\,136 are from
\citet{1996ApJ...466..254B}.}
\label{Table:LMC_clusters}
\begin{tabular}{lccccccc}
\hline 
name & $M$    & $\rcore$ & $\rcore/\rhm$ & age & $\trlx$ & $\trlx^0$ & $\mbh$\\ 
     &$\msun$ & pc &  & Myr & Myr & Myr & $\msun$\\ 
\hline 
R\,136    & 35000 & 0.32 &0.29&  3 &   7--17 &  6--14 & 1000 \\ 
NGC\,2004 & 27000 & 1.57 &0.50& 20 & 71--170 & 41--98 & 1600 \\ 
NGC\,2100 & 30200 & 1.22 &0.62& 16 & 51--120 & 31--73 & 2200 \\ 
\hline
\end{tabular}
\end{table}
The remaining three clusters are listed in
Tab.\,\ref{Table:LMC_clusters}.  For these clusters we reconstruct the
initial relaxation time via the method described by
\citet{2006astro.ph.10659P}.  In this relatively simple model, the
current relaxation time has a one-parameter relation with the
cluster's initial relaxation time.  Of the three clusters listed in
Tab.\,\ref{Table:LMC_clusters}, one (R\,136) is too young to produce
an IMBH and experience a strong encounter with a binary.  The best
candidate clusters to host an IMBH are then NGC\,2100 and
NGC\,2004. We can estimate the mass of a possible black hole in these
clusters by adopting the relation between the cluster structural
parameters and the mass of a central black hole presented by
\citet{hh06}. The authors performed direct N-body simulations to
quantify the relation between the mass of a central IMBH and the ratio
of the core radius to the half-mass radius.  The last column in
Tab.\,\ref{Table:LMC_clusters} provides an estimate for the mass of
the IMBH using this relation. We find that, given the structure of the
observed clusters, a 1600$\msun$ to 2200$\msun$ IMBH could be present.

We argue that NGC\,2100 and NGC\,2004 form the most promising birth
places for \he. If this were the case, the travel time from the parent
cluster to the current location would be less than about 20\,Myr.
Such a short travel time would require an ejection velocity $\apgt
800\kms$ for the adopted total distance to the LMC of 18$\kpc$. This
large velocity could only be obtained with a black hole of several
$10^3\msun$, somewhat higher than the estimated mass of a possible
black hole in these clusters.

We now compute the rate for ejection of hypervelocity stars from
NGC\,2100 and NGC\,2004 using the parameters listed in
Tab.\,\ref{Table:LMC_clusters} and a dimensionless cross-section for
exchange encounters $\tilde{\sigma} \simeq 5000$, as derived from the
scattering experiments.  Adopting a central density of
$2\times10^4\pcc$ (consistent with a King $W_0 = 9$ initial profile),
a velocity of $10\kms$ and a semi-major axis of 0.2\,AU, we obtain an
encounter rate of about one per 2\,Myr for both clusters.  Since only
one in about ten exchange encounters produces a fast escaper (see
Fig.\,\ref{fig:branchHVS}), we derive a production rate of one per
20\,Myr. This value represents a lower limit as we don't take into
account the effects of a mass function in the core and the semi-major
distribution. It is therefore conceivable that NGC\,2100 or NGC\,2004
has produced one high velocity escaper, in which case an IMBH must be
present in one of these clusters.


\section{Acknowledgments}
We are grateful to Jes\'us Ma\'iz Apell\'aniz for interesting comments on
the manuscript.  This work was supported by the Netherlands
Organization for Scientific Research (NWO under grant No. 635.000.001
and 643.200.503), the Royal Netherlands Academy of Arts and Sciences
(KNAW) and the Netherlands Research School for Astronomy (NOVA).

\bibliographystyle{mn2e}
\bibliography{biblio}

\end{document}